\documentclass{aa}
\usepackage{graphicx}
\usepackage{url}
\newcommand{\EQ}{\begin{equation}}
\newcommand{\EN}{\end{equation}}
\newcommand{\EQA}{\begin{eqnarray}}
\newcommand{\ENA}{\end{eqnarray}}
\newcommand{\eq}[1]{(\ref{#1})}

\newcommand{\Eq}[1]{Eq.~(\ref{#1})}
\newcommand{\Eqs}[2]{Eqs~(\ref{#1}) and~(\ref{#2})}

\newcommand{\eqs}[2]{(\ref{#1}) and~(\ref{#2})}

\newcommand{\Sec}[1]{Sect.~\ref{#1}}

\newcommand{\Fig}[1]{Fig.~\ref{#1}}

\newcommand{\bra}[1]{\langle #1\rangle}

\newcommand{\meanF}{\overline{\cal F}} 
\newcommand{\meanEMF}{\overline{\vec{\cal E}}}

\newcommand{\meanFF}{\overline{\mbox{\boldmath ${\cal F}$}} {}}

\newcommand{\meanB}{\overline{B}}

\newcommand{\meanJ}{\overline{J}}
\newcommand{\meanU}{\overline{U}}

\newcommand{\meanAA}{\overline{\vec{A}}}
\newcommand{\meanBB}{\overline{\vec{B}}}
\newcommand{\meanJJ}{\overline{\vec{J}}}
\newcommand{\meanUU}{\overline{\vec{U}}}

\newcommand{\meanEE}{\overline{\vec{E}}}
{}
{}
{}
{}
{}
{}
\newcommand{\UU}{{\vec{U}}}
\newcommand{\uu}{{\vec{u}}}
\newcommand{\BB}{{\vec{B}}}
\newcommand{\JJ}{{\vec{J}}}
\newcommand{\jj}{{\vec{j}}}
\newcommand{\AAA}{{\vec{A}}}

\newcommand{\bb}{{\vec{b}}}
\newcommand{\cc}{{\vec{c}}}

\newcommand{\ee}{\mbox{\boldmath $e$} {}}
\newcommand{\ff}{\mbox{\boldmath $f$} {}}

\newcommand{\EE}{{\vec{E}}}
\newcommand{\FF}{{\vec{F}}}

\newcommand{\nab}{\vec{\nabla}}

\newcommand{\oo}{\vec{\omega}}

\newcommand{\SSSS}{\mbox{\boldmath ${\sf S}$} {}}

\newcommand{\emf}{\mbox{\boldmath ${\cal E}$} {}}

\newcommand{\DD}{{\rm D} {}}

\newcommand{\const}{{\rm const}  {}}

\def\ga{\mathrel{\mathchoice {\vcenter{\offinterlineskip\halign{\hfil
$\displaystyle##$\hfil\cr>\cr\sim\cr}}}
{\vcenter{\offinterlineskip\halign{\hfil$\textstyle##$\hfil\cr>\cr\sim\cr}}}
{\vcenter{\offinterlineskip\halign{\hfil$\scriptstyle##$\hfil\cr>\cr\sim\cr}}}
{\vcenter{\offinterlineskip\halign{\hfil$\scriptscriptstyle##$\hfil\cr>\cr\sim\cr}}}}}
\def\half{{\textstyle{1\over2}}}

\def\onethird{{\textstyle{1\over3}}}

\newcommand{\G}{\,{\rm G}}
\newcommand{\Gtwopers}{\,{\rm G^2\!/s}}

\newcommand{\s}{\,{\rm s}}
\newcommand{\mpers}{\,{\rm m/s}}

\newcommand{\cm}{\,{\rm cm}}

\newcommand{\Mm}{\,{\rm Mm}}
\newcommand{\Mx}{\,{\rm Mx}}

\newcommand{\yjgr}[3]{ #1, {JGR,} {#2}, #3}

\newcommand{\ysph}[3]{ #1, {Sol. Phys.,} {#2}, #3}
\newcommand{\yapj}[3]{ #1, {ApJ,} {#2}, #3}
\newcommand{\yapjl}[3]{ #1, {ApJ,} {#2}, #3}

\newcommand{\yan}[3]{ #1, {AN,} {#2}, #3}

\newcommand{\yana}[3]{ #1, {A\&A,} {#2}, #3}

\newcommand{\ygafd}[3]{ #1, {Geophys. Astrophys. Fluid Dyn.,} {#2}, #3}

\newcommand{\yjfm}[3]{ #1, {JFM,} {#2}, #3}

\newcommand{\yprl}[3]{ #1, {PRL,} {#2}, #3}

\newcommand{\yptrs}[3]{ #1, {Phil. Trans. Roy. Soc.,} {#2}, #3}
\newcommand{\ymn}[3]{ #1, {MNRAS,} {#2}, #3}

\newcommand{\ypre}[3]{ #1, {PRE,} {#2}, #3}

\newcommand{\yjour}[4]{ #1, {#2}, {#3}, #4}

\newcommand{\ybook}[3]{ #1, {#2} (#3)}

\newcommand{\sjour}[2]{ #1, {#2} (submitted)}
\begin{document}

\title{Catastrophic alpha quenching alleviated by helicity flux and shear}
\author{Axel Brandenburg\inst{1} \and Christer Sandin\inst{2}}

\institute{
NORDITA, Blegdamsvej 17, DK-2100 Copenhagen \O, Denmark
\and
Astrophysical Institute  Potsdam, An der Sternwarte 16, D-14482 Potsdam, Germany
}

\date{
Received 16 January 2004 / accepted 21 July 2004}

\abstract{
A new simulation set-up is proposed for studying mean field dynamo action.
The model combines the computational advantages of local cartesian geometry
with the ability to include a shear profile that resembles the sun's
differential rotation at low latitudes.
It is shown that in a two-dimensional mean field model this geometry
produces cyclic solutions with dynamo waves traveling away from
the equator -- as expected for a positive alpha effect in the northern
hemisphere.
In three dimensions with turbulence driven by a helical forcing function,
an alpha effect is self-consistently generated in the presence of a
finite imposed toroidal magnetic field.
The results suggest that, due to a finite flux of current helicity out
of the domain, alpha quenching appears to be non-catastrophic -- at least
for intermediate values of the magnetic Reynolds number.
For larger values of the magnetic Reynolds number, however, there is
evidence for a reversal of the trend and that $\alpha$ may decrease
with increasing magnetic Reynolds number.
Control experiments with closed boundaries confirm that in the
absence of a current helicity flux, but with shear as before,
alpha quenching is always catastrophic and
alpha decreases inversely proportional to
the magnetic Reynolds number.
For solar parameters, our results suggest a current helicity flux of
about $0.001\Gtwopers$.
This corresponds to a magnetic helicity flux, integrated over the northern
hemisphere and over the 11 year solar cycle, of about
$10^{46}\Mx^2$.
}

\offprints{CSandin@aip.de}

\maketitle

\section{Introduction}

The large scale magnetic field of stars and galaxies is often interpreted
in terms of mean field dynamo theory, which
takes into account that the turbulence is at least partially
helical.
The helicity, in turn, leads to the so-called $\alpha$ effect, i.e.\ an averaged
field-aligned current that induces new field loops perpendicular to the
original field (Moffatt 1978, Krause \& R\"adler 1980).
In the presence of shear, poloidal loops get sheared out, thereby reinforcing
the toroidal field from which more poloidal loops can be created.
This is the basic $\alpha\Omega$ dynamo mechanism which is at least in
principle able to explain the cyclic variation of the solar magnetic field
(Parker 1979).

The investigation of mean field dynamos
has been the subject of numerous papers since the 1970s
(see also Zeldovich et al.\ 1983).
With the advent of high resolution turbulence simulations it has become
exceedingly clear that there is a serious problem in the nonlinear
case at large magnetic Reynolds numbers.
Similar problems are completely unknown in the context of nonmagnetic,
purely hydrodynamic
turbulence or in nonhelical hydromagnetic turbulence where the associated
dissipative fluxes always remain finite. 
This is not the case with the magnetic helicity flux which goes to zero
in the large magnetic Reynolds number limit (Berger 1984).
The magnetic helicity is thus almost perfectly conserved in
practically all astrophysically interesting cases.

An important consequence of magnetic helicity conservation is the fact that
the $\alpha$ effect cannot produce any net magnetic helicity.
This means that if the $\alpha$ effect produces large scale magnetic fields,
it must at the same time also give rise to a certain amount of
small scale fields with opposite sign of magnetic helicity
(Seehafer 1996, Ji 1999).
Hence the strength of the large scale field that can be generated on
dynamical time scales is limited, as the associated
small scale field cannot grow significantly above the equipartition
field strength (Brandenburg 2001, hereafter B01, Field \& Blackman 2002,
Blackman \& Brandenburg 2002, Subramanian 2002).
Recent work has shown that this leads to a rather restrictive nonlinearity
of the $\alpha$ effect.
It has been recognized for some time (Blackman \& Field 2000a,b,
Kleeorin et al.\ 2000) that the conservation of magnetic
helicity may be particularly restrictive in the presence of closed or
periodic boundary conditions used in many investigations.
While it is clear that magnetic helicity fluxes through boundaries
can help in principle (Brandenburg et al.\ 2002), one must still ensure
that it also is of the right properties. If the
magnetic helicity flux carries away most of the desired large scale field,
nothing will be gained and the dynamo will be even less efficient.
This is indeed what early simulations have shown (Brandenburg \& Dobler 2001).

The purpose of the present paper is to show that the situation changes
considerably when helicity flux is mediated by
shear, the only mechanism known that can separate
and hence also transport magnetic helicity in space (the $\alpha$ effect,
by comparison, separates and transports magnetic helicity in wavenumber space).

\begin{figure}
\resizebox{\hsize}{!}{\includegraphics{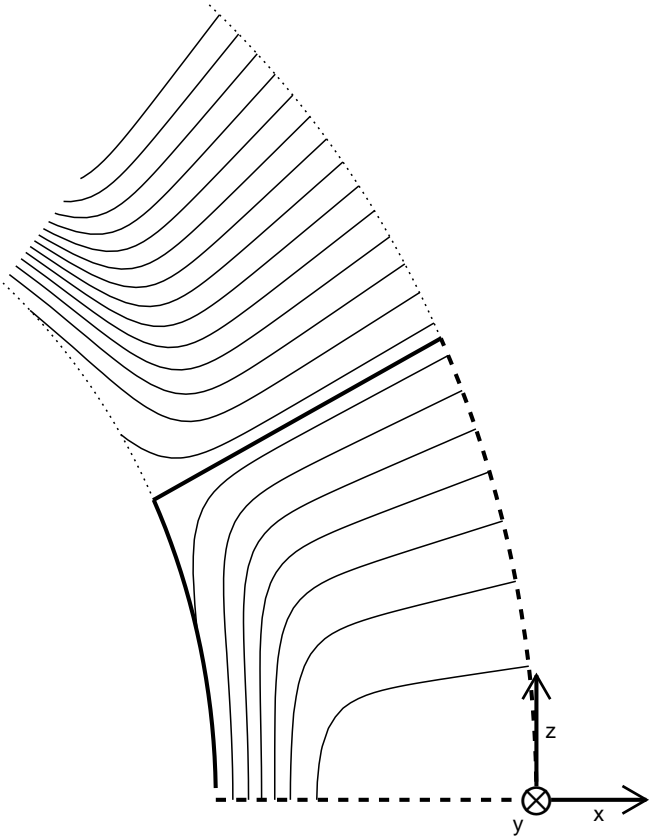}}\caption{
A sketch of the solar angular velocity at low latitudes with
spoke-like contours in the bulk of the convection zone merging
gradually into uniform rotation in the radiative interior.
The low latitude region, modeled in this paper, is
indicated by thick lines.
}\label{sketch1}\end{figure}

We adopt the simplest possible model that is able to capture the effects
of helicity transport from one hemisphere to the other.
To motivate our model we first look at an idealized representation of the
solar angular velocity which is spoke-like in the bulk of the convection
zone and nearly rigid in the radiative interior; see \Fig{sketch1}.

\begin{figure}[t!]
\centering\includegraphics[width=0.5\textwidth]{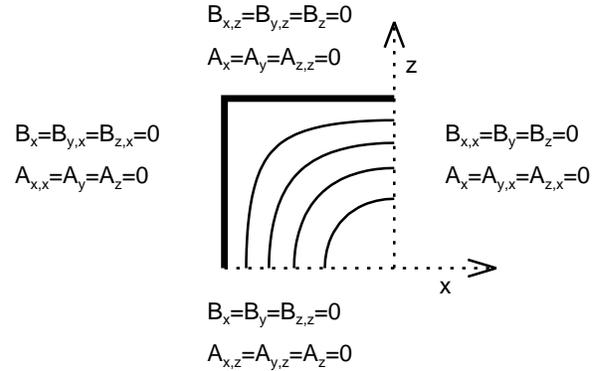}\caption{
Differential rotation in our cartesian model, with
the equator being at the bottom, the surface to the right,
the bottom of the convection zone to the left and mid-latitudes
at the top.
The boundary conditions for the three components of the magnetic field
and the vector potential (indicated near the boundaries of
the box) are discussed in \Sec{Sapproach}.
}\label{sketch2}\end{figure}

We model the region below $30^\circ$ latitude
by adopting a cartesian geometry where
the $x$ direction corresponds to radius, the $y$ direction to longitude,
and the $z$ direction to latitude.
We ignore the fact that in the sun the radial transition to
uniform rotation is much steeper and
model the mean toroidal velocity simply in terms of trigonometric functions
using
\begin{equation}
\meanUU=U_0\cos k_1x\cos k_1z,
\end{equation}
where $k_1$ is the lowest wavenumber in the $(x,z)$ plane with
$-\pi/2\leq k_1x\leq0$ and $0\leq k_1z\leq\pi/2$.
In the following we adopt units where $k_1=1$.
The equator is assumed to be at $z=0$ and the outer surface at $x=0$.
The bottom of the convection zone is at $x=-\pi/2$ and the latitude where
the surface angular velocity equals the value in the radiative interior
is at $z=\pi/2$; see \Fig{sketch2}.

In order to clarify some basic properties of this rather unexplored
geometry we begin by studying $\alpha\Omega$ dynamos in this geometry
in \Sec{Smeanfield}, after introducing our numerical approach
in \Sec{Sapproach}.
The problem is at first reduced to two dimensions.
Here we assume that there is an $\alpha$ effect so that a mean magnetic
field can readily be generated by this term.
Next we drive helical turbulence by a corresponding forcing term in
three dimensions and measure the resulting $\alpha$ effect by imposing
a toroidal magnetic field in \Sec{TurbulenceSimulations}.
In particular, we study the dependence of $\alpha$ quenching on the
magnetic Reynolds number and show that the $\alpha$ effect is only
catastrophically quenched when the boundaries are closed or when the
large scale shear has been turned off.

\section{Our numerical approach}
\label{Sapproach}

The evolution of the magnetic field, $\BB$, is governed by the induction
equation, $\partial\BB/\partial t=-\nab\times\EE$, where $\EE$ is the
electric field.
The induction equation is solved in terms of the magnetic vector potential,
$\AAA$, where $\BB=\nab\times\AAA$.
The evolution of $\AAA$ is therefore given by
$\partial\AAA/\partial t=-\EE-\nab\phi$, where $\phi$ is the scalar
potential.
By using the gauge in which $\phi=0$ we simply have
\begin{equation}
\partial\AAA/\partial t=-\EE,
\label{induction}
\end{equation}
which is the equation that will be considered throughout.
The (negative) electric field is given by
\begin{equation}
-\EE=\UU\times\BB-\eta\mu_0\JJ,
\end{equation}
where $\UU$ is the velocity, $\JJ=\nab\times\BB/\mu_0$ the current density,
and $\mu_0$ the vacuum permeability.

At the bottom of the convection zone and at mid-latitudes we assume
the same boundary conditions for the magnetic field as for the velocity
field, i.e.\ the field is tangent to the boundaries, which corresponds to
perfect conductor boundary conditions.
This means that the tangential electric field vanishes, and, because
of the gauge $\phi=0$, we have
\begin{equation}
A_x=A_y=0\quad\mbox{on $z=\pi/2$},\quad\mbox{and}
\end{equation}
\begin{equation}
A_y=A_z=0\quad\mbox{on $x=-\pi/2$}.
\end{equation}
On the equator and at the outer surface we assume that the magnetic field
is normal to the boundaries, i.e.\
\begin{equation}
B_x=B_y=0\quad\mbox{on $z=0$},\quad\mbox{and}
\label{open1}
\end{equation}
\begin{equation}
B_y=B_z=0\quad\mbox{on $x=0$}.
\label{open2}
\end{equation}
At the equator this boundary condition is consistent with dipolar
parity of the field.
The full set of boundary conditions for all three components of both
$\AAA$ and $\BB$ are given on the four `meridional' boundaries of the
box in \Fig{sketch2}.
In the $y$ direction we adopt periodic boundary conditions over the
interval $0<y<2\pi$.

Occasionally we refer to the boundary conditions \eqs{open1}{open2}
as open, because
they permit flux of magnetic and current helicities through the
$z=0$ and $x=0$ boundaries.
For comparison we also perform calculations with closed boundaries
where $A_x=A_y=0$ on $z=0$ and $A_y=A_z=0$ on $x=0$.
With these conditions the magnetic and current helicity fluxes vanish
through these boundaries.
At the bottom of the convection zone and at mid-latitudes the fluxes
of magnetic and current helicities are always vanishing.
This is probably a reasonable assumption, because at these locations
there is no shear to mediate large scale helicity transport and the
small scale helicity transport was already previously found to fluctuate
around zero if there is no shear (Brandenburg \& Dobler 2001).

For both the mean field calculations and the turbulence simulations we
step the equations forward in time by using the Pencil Code\footnote{
\url{http://www.nordita.dk/software/pencil-code}}.
For the mean field calculations a typical resolution of $32^2$ meshpoints
proved to be sufficient, while for the turbulence simulations the
required resolution depends on the magnetic Reynolds number, $R_{\rm m}$.
Here, $R_{\rm m}$ is based on the magnitude of the turbulent velocity
and not on the larger shear flow velocity.
(The precise definition is given below in \Sec{ResultsAlphaQuenching}.)
For $R_{\rm m}\approx100$, a resolution of $512^3$ meshpoints is required,
while for $R_{\rm m}\approx15$, a resolution of $128^3$ meshpoints proved
to be sufficient.
We note, however, that the aspect ratio of the box is 1:4:1 and, although
there is shear smearing out structures in the $y$ direction, a uniform
mesh aspect ratio seems often to be preferred.
For example, a run with $128\times512\times128$ meshpoints allowed us
to use higher Reynolds numbers than $128^3$ meshpoints.

\section{Mean field calculations}
\label{Smeanfield}

An important aspect of our studies is to show that the shear flow
depicted in \Fig{sketch2} is a reasonable approximation to the differential
rotation present in the sun (which is more like that depicted
in \Fig{sketch1}).

In the context of mean field theory it is known that in spherical shells
both dipolar and quadrupolar solutions are approximately equally easily
excited (Roberts 1972)
and that both solutions can be oscillatory with field migration
away from the midplane (when $\alpha>0$).
We want to know whether in the present geometry the magnetic field
evolution is similar to that in spherical shells.

The mean field induction equation is given by
$\partial\meanBB/\partial t=-\nab\times\meanEE$,
which again is solved in terms of the mean magnetic vector potential,
i.e.\ $\partial\meanAA/\partial t=-\meanEE$, where
\begin{equation}
-\meanEE=\meanUU\times\meanBB+\meanEMF-\eta\mu_0\meanJJ.
\end{equation}
Here, overbars denote azimuthal averages, and velocity and magnetic field
are split into mean and fluctuating components via
$\BB=\meanBB+\bb$ and $\UU=\meanUU+\uu$.
The electromotive force from the fluctuating components of velocity
and magnetic field, $\meanEMF=\overline{\uu\times\bb}$, is in its
simplest form (e.g.\ Moffatt 1978, Krause \& R\"adler 1980)
\begin{equation}
\meanEMF=\alpha\meanBB-\eta_{\rm t}\mu_0\meanJJ,
\end{equation}
where $\alpha$ (related to the mean helicity)
and $\eta_{\rm t}$ (turbulent diffusivity) could still
be functions of $x$, $z$, and $t$, as well as a function of $\meanBB$
itself, but for simplicity we assume them to be constant here.

The solutions are characterized by two non-dimensional parameters,
\begin{equation}
C_\alpha=\alpha/(\eta_{\rm T}k_1),\quad\mbox{and}\quad
C_S=U_0/(\eta_{\rm T}k_1),\label{CaCs}
\end{equation}
where $\eta_{\rm T}=\eta+\eta_{\rm t}$ is the sum of microscopic and
turbulent magnetic diffusivities.

\begin{figure}
\resizebox{\hsize}{!}{\includegraphics{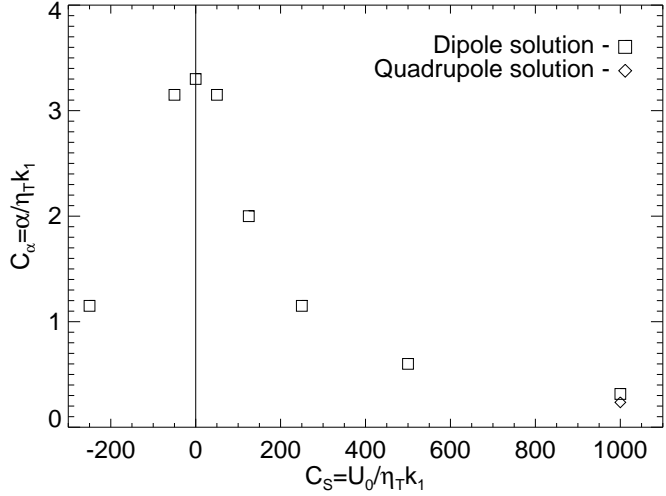}}\caption{
Critical value of $C_\alpha$ for dynamo action as a function of
shear parameter, $C_S$.
Note the typical decrease of the critical value of $C_\alpha$ with
increasing $C_S$.
}\label{growth}\end{figure}

In \Fig{growth} we plot the stability diagram in the $(C_\alpha,C_S)$ plane.
For $C_\alpha<C_{\alpha,{\rm crit}}$ the solutions are decaying and for
$C_\alpha>C_{\alpha,{\rm crit}}$ they are growing exponentially and are
oscillatory (Hopf bifurcation), except for a narrow interval around
$C_S=0$. 
Such a behavior is quite typical of $\alpha\Omega$ dynamos (see, e.g.,
Roberts \& Stix 1972).
For $C_S=1000$ we have also considered the quadrupolar solution and
find that it is slightly easier to excite (see \Fig{growth}).
As stated earlier, the approximately equal excitation conditions for
dipolar and quadrupolar solutions, seen in \Fig{growth},
is typical of $\alpha\Omega$ dynamos in spherical shells.
Indeed, the fact that quadrupolar solutions can be preferred has been
found in other solar dynamo models (Dikpati \& Gilman 2001).

\begin{figure}
\resizebox{\hsize}{!}{\includegraphics{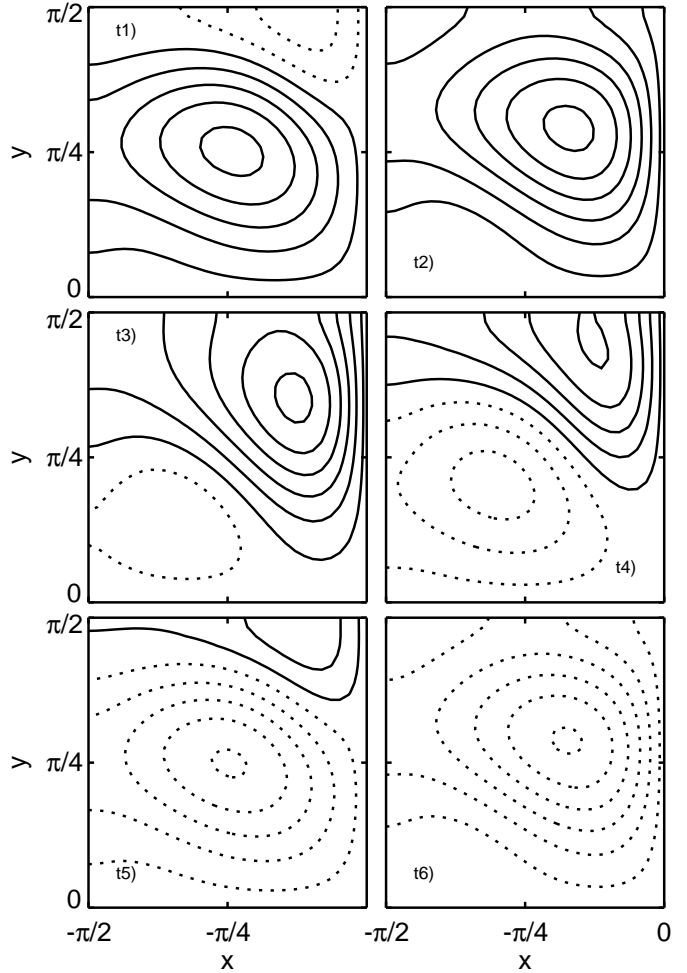}}\caption{
Contours of $\meanB_y$ for times separated by
$\Delta t=300$ in the temporal sequence t1-t6.
Note the field migration from the lower left corner to the upper right;
in this figure $C_S=1000$ [see Eq. (\ref{CaCs})].
}\label{pslices}\end{figure}

In \Fig{pslices} we show contours of $\meanB_y$ for the marginally excited case
with $C_S=1000$ for different times covering a little more than half a cycle.
One clearly sees magnetic field migration away from the equator.
We note that in the sun the field migration is toward the equator.
The reason for this is not entirely clear, but it could be caused by
a negative $\alpha$ effect (but the reason for this is not clear either)
or by meridional circulation (Choudhuri et al.\ 1995, Durney 1995).
However, before these questions can seriously be addressed, it is
mandatory to have a reliable mean field theory.
This has so far been hampered by not being able to model the nonlinear
feedback correctly.

Based on the similarity of the stability diagram and the migration pattern
in calculations using cartesian geometry and spherical shells
(e.g.\ Roberts \& Stix 1972),
we may conclude that our local model provides a reasonable approximation
to the more realistic case of a spherical shell.
Since azimuthal averages are used, the mean field equations would be
axisymmetric and could easily be solved using meshpoint methods.
The case of three-dimensional turbulence in spherical shells is considerably
more difficult because the coordinate singularity at the poles
leads to serious timestep restrictions in the azimuthal direction.

\section{Turbulence simulations}
\label{TurbulenceSimulations}

In this section we consider three-dimensional turbulence in the same
cartesian geometry that we used in the previous section.
We now consider the full (non-averaged) velocity and magnetic fields,
$\UU$ and $\BB$, respectively.

\subsection{Basic equations}

The fluctuating velocity together with the shear flow must be
obtained by simultaneously solving
the induction equation [Eq.\ (\ref{induction})] together with the
momentum equation, which we write here for an isothermal gas
of constant sound speed $c_{\rm s}$,
\begin{equation}
{\DD\UU\over\DD t}=-c_{\rm s}^2\nab\ln\rho+{\JJ\times\BB\over\rho}
+\ff+\FF_{\rm visc},
\end{equation}
where $\ff$ is the forcing function driving both the turbulence
(around a narrow band of wavenumbers around $k_{\rm f}=5$) and the
shear flow (cf.\ Brandenburg et al.\ 2001). Moreover,
\begin{equation}
\FF_{\rm visc}=\nu\left(\nabla^2\UU+\onethird\nab\nab\cdot\UU
+2\SSSS\cdot\nab\ln\rho\right)
\end{equation}
is the viscous force where $\nu=\mbox{const}$ is the kinematic viscosity,
${\sf S}_{ij}=\frac{1}{2}(U_{i,j}+U_{j,i})-\frac{1}{3}\delta_{ij}U_{k,k}$
the traceless rate of strain tensor, and $\rho$ the density which
obeys the continuity equation which we solve in the form
\begin{equation}
{\DD\ln\rho\over\DD t}=-\nab\cdot\UU,
\end{equation}
where $\DD/\DD t=\partial/\partial t+\UU\cdot\nab$ is the advective
derivative.
We adopt a random forcing function with finite helicity; see
B01 for details.

\subsection{Results for $\alpha$ quenching}
\label{ResultsAlphaQuenching}

\begin{figure}\centering
\resizebox{.92\hsize}{!}{\includegraphics{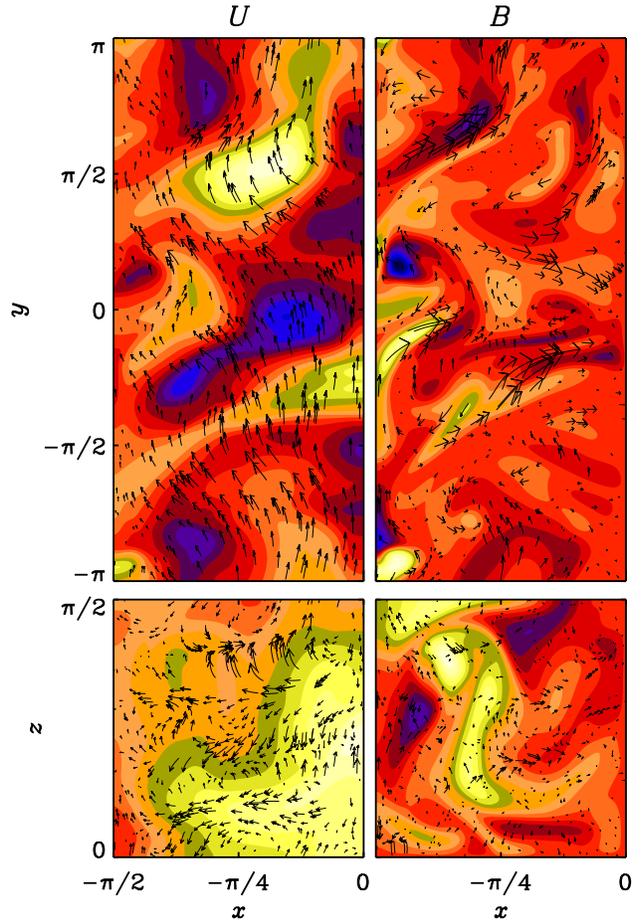}}\caption{
Vectors of $\UU$ and $\BB$ in an $xy$ plane through $z=\pi/4$,
superimposed on a grey/color representation of their normal components,
for a run with $R_{\rm m}=14$, open boundaries, shear, and
negative helicity.
Note that the velocity field is dominated by the toroidal shear flow.
}\label{pslice}\end{figure}

We have carried out a range of simulations for different values of the
magnetic Reynolds number,
\EQ
R_{\rm m}=u_{\rm rms}/(\eta k_{\rm f}),
\EN
for both open and closed boundary conditions.
(Here, $u_{\rm rms}$ does not include the mean shear flow.)
In order to measure $\alpha$, a uniform magnetic field,
$\BB_0=\const$, is imposed, and the magnetic field is now written as
$\BB=\BB_0+\nab\times\AAA$.
In all cases presented below we have used
$B_0\approx0.07\sqrt{\mu_0\rho}\,u_{\rm rms}$.
In \Fig{pslice} we show a
graphical presentation of a typical snapshot of a solution.

We have determined $\alpha$ by measuring the turbulent
electromotive force, i.e.\ $\alpha=\bra{\emf}\cdot\BB_0/B_0^2$.
Similar investigations have been done before both for forced turbulence
(e.g., Cattaneo \& Hughes 1996, see also B01) and for convective turbulence
(e.g., Brandenburg et al.\ 1990, Ossendrijver et al.\ 2001).
The contribution from $\eta_{\rm T}\meanJJ$ has been ignored in this
approach; this is justified because for strong shear the poloidal field,
giving rise to a toroidal mean current, is weak.

\begin{figure}[t!]
\centering\includegraphics[width=0.5\textwidth]{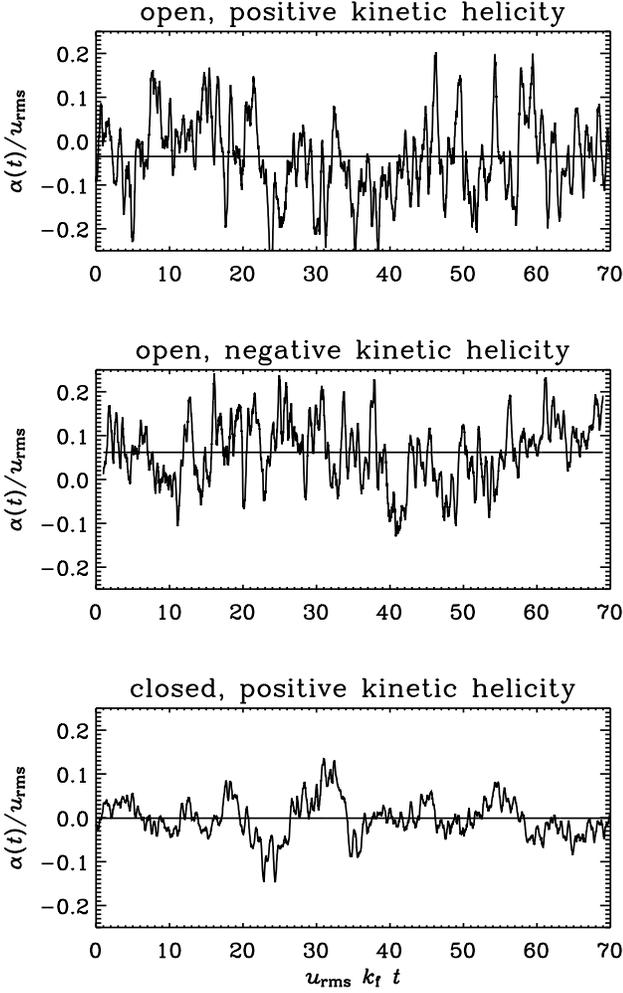}\caption{
Time traces of $\alpha(t)$ for runs with open boundaries
(for both positive and negative kinetic helicity) and
closed boundaries (for positive kinetic helicity).
$R_{\rm m}=30$ in all cases.
}\label{palp_hires}\end{figure}

It is well known that the $\alpha$ effect is an extremely noisy quantity
-- especially in the case of large magnetic Reynolds numbers
(Cattaneo \& Hughes 1996).
The strong fluctuations are also clear from \Fig{palp_hires},
where we plot $\alpha(t)$ both for open and closed boundaries.
Note that $\alpha/u_{\rm rms}$ fluctuates in time in the range $\pm0.2$
about a much smaller average value of, e.g., $-0.03$ in the top panel.

As expected, $\alpha$ is negative when the helicity of the forcing is
positive, and $\alpha$ changes sign when the helicity of the forcing
changes sign.
For $R_{\rm m}\ga14$ 
the magnitudes of $\alpha$ begin to be different in the two cases:
$|\alpha|$ is larger when the helicity of the forcing is negative.
In the sun, this corresponds to the sign of helicity in the northern
hemisphere in the upper parts of the convection zone.
This is here the relevant case, because the differential rotation
pattern of our model also corresponds to the northern hemisphere.

\begin{figure}
\resizebox{\hsize}{!}{\includegraphics{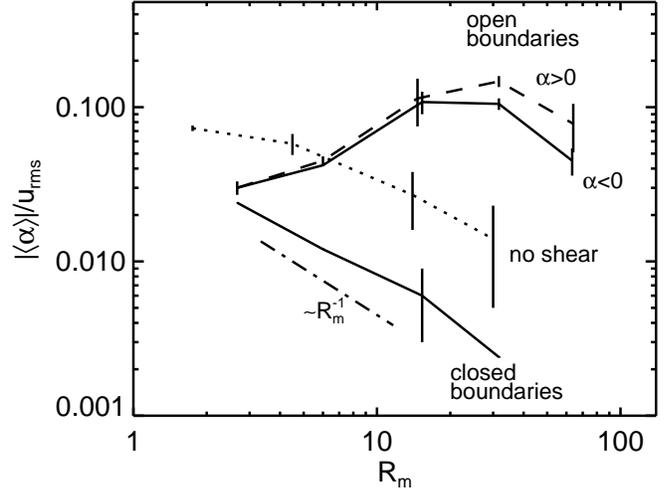}}\caption{
Dependence of $|\bra{\alpha}|/u_{\rm rms}$ on $R_{\rm m}$
for open and closed boundaries.
The case with open boundaries and negative helicity is shown as a dashed line.
Note that for $R_{\rm m}\approx30$ the $\alpha$ effect
is about 30 times smaller when the boundaries are closed.
The dotted line gives the result with open boundaries but no shear.
The vertical lines indicate the range obtained by calculating
$\alpha$ using only the first and second half of the time interval.
}\label{palp_sum}\end{figure}

There is a striking difference between the cases with open and
closed boundaries which becomes particularly clear when comparing
the averaged values of $\alpha$ for different magnetic Reynolds
numbers; see \Fig{palp_sum}.
With closed boundaries $\alpha$ tends to zero like $R_{\rm m}^{-1}$,
while with open boundaries $\alpha$ shows no such immediate decline;
only for larger values of $R_{\rm m}$ there is possibly an asymptotic
$\alpha\propto R_{\rm m}^{-1}$ dependence.
There is also a clear difference between the cases with and without shear.
In the absence of shear (dotted line in \Fig{palp_sum}) $\alpha$ declines
with increasing $R_{\rm m}$, even though for small values of $R_{\rm m}$
it is larger than with shear.
This suggests that the presence of shear combined with open boundaries
might be a crucial prerequisite of dynamos that saturate on a dynamical
time scale.

The difference between open and closed boundaries
will now be discussed in terms of a current helicity
flux through the two open open boundaries of the domain.

\subsection{Current helicity flux}

It is suggestive to interpret the above results in terms of the
dynamical $\alpha$ quenching model, where $\alpha$ is proportional to the
difference between kinetic and current helicities (Pouquet et al.\ 1976), i.e.\
\begin{equation}
\alpha=-\onethird\tau\left(
\overline{\oo\cdot\uu}-\rho_0^{-1}\overline{\jj\cdot\bb}\right)
\equiv\alpha_{\rm K}+\alpha_{\rm M}.
\end{equation}
Here, $\oo=\nab\times\uu$ is the small scale vorticity,
$\jj=\nab\times\bb/\mu_0$ is the small scale current density,
$\alpha_{\rm K}$ is the kinematic
contribution to the $\alpha$ effect, and $\alpha_{\rm M}$ the
magnetic contribution primarily responsible for the quenching of the
$\alpha$ effect.

In order to obtain an expression for $\alpha_{\rm M}$ we begin
with the evolution equation for $\overline{\jj\cdot\bb}$.
In the absence of boundary conditions it has been advantageous to
start with the evolution equation for magnetic helicity (because
it is conserved) instead of current helicity (which is not conserved);
see Kleeorin \& Ruzmaikin (1982), Kleeorin et al.\ (1995),
Blackman \& Brandenburg (2002).
In the case of open boundary conditions, this approach becomes
cumbersome, because one has to consider the gauge-invariant relative
magnetic helicity (Berger \& Field 1984).
Furthermore, the concept of a {\it density} of magnetic helicity is
not meaningful, because it would depend on the gauge.
In order to avoid these problems it is advantageous consider the
current helicity equation (Subramanian \& Brandenburg 2004,
Brandenburg \& Subramanian 2004).
Apart from this technicality, the following derivation is similar to
that of Blackman \& Brandenburg (2002) for the case without current
helicity flux.

Using the evolution equation, $\partial\bb/\partial t=-\nab\times\ee$,
for the fluctuating magnetic field, where $\ee=\EE-\meanEE$ is the
small scale electric field and $\meanEE=\eta\meanJJ-\meanEMF$ the
mean electric field, we can derive the equation
\begin{equation}
{\partial\over\partial t}\overline{\jj\cdot\bb}
=-2\,\overline{\ee\cdot\cc}-\nab\cdot\meanFF_C^{\rm SS},
\label{jc_evolution}
\end{equation}
where
\begin{equation}
\meanFF_C^{\rm SS}=\overline{2\ee\times\jj}+\overline{(\nab\times\ee)\times\bb}
\label{meanFFc}
\end{equation}
is the current helicity flux from the small scale field, and
$\cc=\nab\times\jj$ the curl of the small scale current density,
$\jj=\JJ-\meanJJ$.
In the isotropic case,
$\overline{\ee\cdot\cc}\approx k_{\rm f}^2\overline{\ee\cdot\bb}$, where
$k_{\rm f}$ is the typical wavenumber of the fluctuations,
here assumed to be the forcing wavenumber.
Ignoring the effect of the mean flow on $\meanEMF$ [as is usually done;
but see Krause \& R\"adler (1980) or Rogachevskii \& Kleeorin 2003], we obtain
\begin{equation}
\overline{\ee\cdot\bb}
\approx-\overline{(\uu\times\BB_0)\cdot\bb}+\eta\overline{\jj\cdot\bb}
=\meanEMF\cdot\meanBB+\eta\overline{\jj\cdot\bb},
\end{equation}
where we have used $\overline{\uu\times\bb}=\meanEMF$ and
$\BB_0=\meanBB$.
Using standard expressions for the turbulent magnetic diffusivity,
$\eta_{\rm t}=\onethird\tau u_{\rm rms}^2$, and the equipartition
field strength, $B_{\rm eq}=\sqrt{\mu_0\rho}\,u_{\rm rms}$,
we eliminate $\tau$ via
\begin{equation}
\onethird\tau\rho_0^{-1}=\eta_{\rm t}/B_{\rm eq}^2.
\end{equation}
This leads to an explicitly time dependent formula for $\alpha$,
\begin{equation}
{\partial\alpha\over\partial t}=-2\eta_{\rm t} k_{\rm f}^2\left(
{\meanEMF\cdot\meanBB
+\half k_{\rm f}^{-2}\nab\cdot\meanFF_C^{\rm SS} \over B_{\rm eq}^2}
+{\alpha-\alpha_{\rm K}\over R_{\rm m}}\right).
\label{fullset2flux}
\end{equation}
This equation is similar to that of Kleeorin et al.\ (2000, 2002, 2003)
who considered the flux of magnetic helicity instead of current helicity.

Making the adiabatic approximation, i.e.\ putting the rhs of
\Eq{fullset2flux} to zero, one arrives at the algebraic
steady state quenching formula ($\partial\alpha/\partial t=0$)
\begin{equation}
\alpha={\alpha_{\rm K}
+R_{\rm m}\left(\eta_{\rm t}\mu_0\meanJJ\cdot\meanBB
-\half k_{\rm f}^{-2}\nab\cdot\meanFF_C^{\rm SS}\right)/B_{\rm eq}^2
\over1+R_{\rm m}\meanBB^2/B_{\rm eq}^2}.
\label{AlphaStationaryFlux}
\end{equation}
Furthermore, if the mean field is defined
as an average over the whole box, then $\meanBB\equiv\BB_0=\const$,
so $\meanJJ=0$ and \Eq{AlphaStationaryFlux} reduces to
\begin{equation}
\alpha={\alpha_{\rm K}
-\half k_{\rm f}^{-2}R_{\rm m}\nab\cdot\meanFF_C^{\rm SS}/B_{\rm eq}^2
\over1+R_{\rm m}\BB_0^2/B_{\rm eq}^2}.
\label{AlphaStationaryFlux_noJ}
\end{equation}
This expression applies to the present case, because we consider
only the statistically steady state and we also define the mean field
as a volume average.

For closed boundaries, $\nab\cdot\meanFF_C^{\rm SS}=0$, and so
\Eq{AlphaStationaryFlux_noJ} clearly reduces to a catastrophic quenching
formula, i.e.\ $\alpha$ vanishes in the limit of large magnetic Reynolds
numbers as
\begin{equation}
\alpha^{\rm(closed)}={\alpha_{\rm K}
\over1+R_{\rm m}\BB_0^2/B_{\rm eq}^2}\to R_{\rm m}^{-1}
\quad\mbox{(for $R_{\rm m}\to\infty$)}.
\label{AlphaStationary_noFlux}
\end{equation}
The $R_{\rm m}^{-1}$ dependence suggested by \Eq{AlphaStationary_noFlux}
is confirmed by the simulations
(compare with the dash-dotted line in \Fig{palp_sum}).
On the other hand, for open boundaries the limit $R_{\rm m}\to\infty$
gives
\begin{equation}
\alpha^{\rm(open)}\to
-(\nab\cdot\meanFF_C^{\rm SS})/(2k_{\rm f}^2\BB_0^2)
\quad\mbox{(for $R_{\rm m}\to\infty$)},
\label{AlphaOpenLimit}
\end{equation}
which shows that losses of negative helicity, as observed in the northern
hemisphere of the sun, would enhance a positive $\alpha$ effect
(Kleeorin et al.\ 2000).
In the simulations, the current helicity flux is found to be independent
of the magnetic Reynolds number.
This explains why the $\alpha$ effect no longer shows the catastrophic
$R_{\rm m}^{-1}$ dependence (see \Fig{palp_sum}).

\subsection{Estimates for the Vishniac-Cho flux}

\begin{figure}
\resizebox{\hsize}{!}{\includegraphics{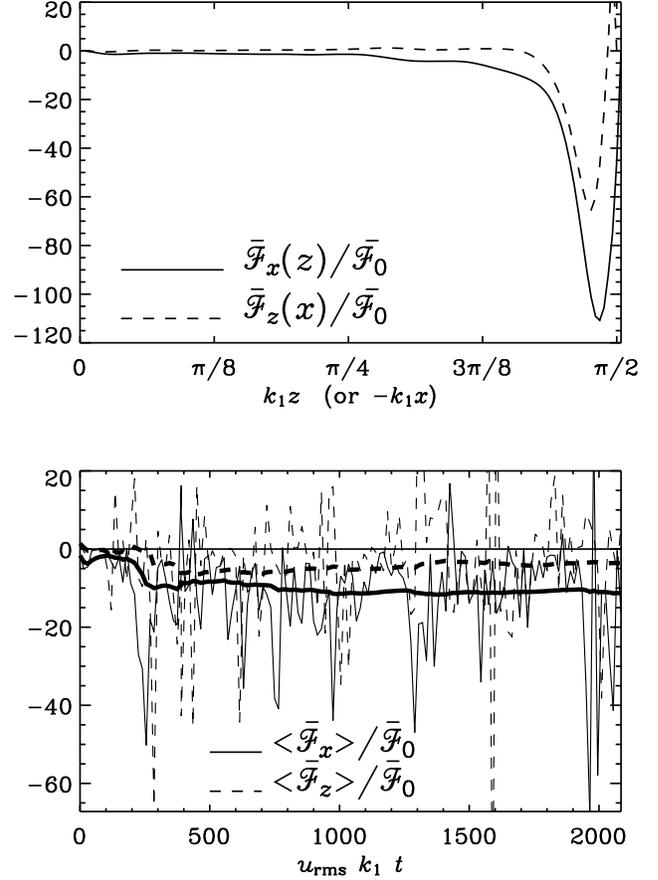}}\caption{
The Vishniac-Cho flux on the outer surface
($\meanF_x^{\rm VC}$) and at the equator ($\meanF_z^{\rm VC}$),
averaged over the $y$
direction, for the run with $\alpha>0$ and $R_{\rm m}=14$.
In the upper panel the flux is also averaged over $t$ and shown as
a function of $z$ and $x$, respectively.
In the lower panel, the $x$ and $z$ components of the fluxes are
averaged over the $z$ and $x$ directions, respectively, and
shown as functions of $t$.
The fat solid and dashed lines denote the running means for the
two functions.
}\label{pvishcho_fin}\end{figure}

Theoretical estimates for magnetic helicity fluxes have been proposed
by Kleeorin et al.\ (2000) and Vishniac \& Cho (2001).
The two fluxes are rather different.
The expression of Vishniac \& Cho (2001) has been confirmed independently
and can be written in the form (Subramanian \& Brandenburg 2004; see also
the review by Brandenburg \& Subramanian 2004)
\EQ
\meanFF_k^{\rm VC}=-4\tau\overline{\omega_i\nabla_j u_k}\;\meanB_i\meanB_j.
\EN
In \Fig{pvishcho_fin} we plot the profiles of this flux, averaged in the
$y$ direction, on the two open boundaries.
We also show the time evolution
of the averaged fluxes on the two open boundaries.
It turns out that the magnitude of the two fluxes is large compared with
$\meanFF_0\equiv u_{\rm rms}k_{\rm f}B_0^2$,
but the fluxes also fluctuate strongly in time, so it is important
to average over long times.
Furthermore, there is a clear tendency for the difference between
incoming flux at the equator ($\meanFF_z^{\rm VC}$, dashed line) and outgoing
fluxes at outer surface ($\meanFF_x^{\rm VC}$, solid line) to cancel partially,
giving a smaller net flux.
Nevertheless, since $|\meanFF_x^{\rm VC}|>|\meanFF_z^{\rm VC}|$,
the net outgoing flux is negative, as expected for the northern hemisphere.

\subsection{Large scale current helicity flux}

\begin{figure}
\resizebox{\hsize}{!}{\includegraphics{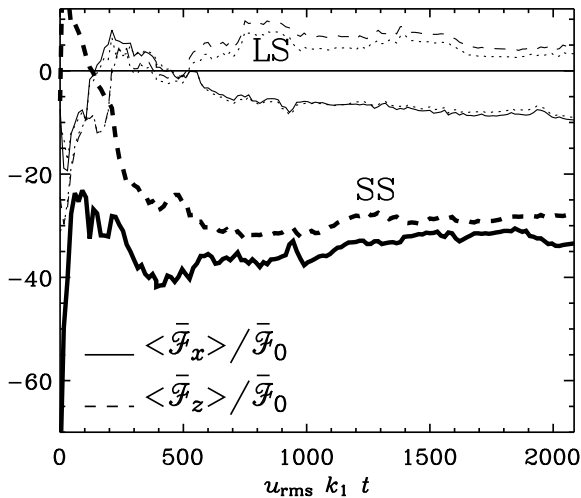}}\caption{
Normal components of the current helicity flux on the outer surface
($\meanF_x$) and at the equator ($\meanF_z$), averaged over the
corresponding surfaces, for the same run as in \Fig{pvishcho_fin}.
The fat lines denote the fluxes from the small
scale field, $\meanFF_C^{\rm SS}$, while the thin lines denote
the fluxes form the large scale field, $\meanFF_C^{\rm LS}$.
The dotted lines near the two $\meanFF_C^{\rm LS}$ curves show
the result of the approximation \eq{meanFFcLS2}.
}\label{pcurhelflux}\end{figure}

In earlier work (Brandenburg \& Dobler 2001) it was reported that the
contribution to the magnetic helicity flux was outweighed by a much
larger flux from the large scale field.
In the present paper we work instead with the current helicity, and the
current helicity flux from the large scale field is
\begin{equation}
\meanFF_C^{\rm LS}=2\meanEE\times\meanJJ
+(\nab\times\meanEE)\times\meanBB.
\label{meanFFcLS}
\end{equation}
For the vertical field condition, see \Eqs{open1}{open2}, the second term
in \Eq{meanFFcLS} vanishes.
Assuming isotropy, the contribution from the first term involves
$\meanEMF=\alpha\meanBB-\eta_{\rm t}\meanJJ$, but this does not contribute
either, because $\alpha\meanBB\times\meanJJ$ does not have a normal
component on the boundaries, and $\eta_{\rm t}\meanJJ\times\meanJJ=0$.
In the first term only the mean flow term contributes, so we get
\begin{equation}
\meanFF_C^{\rm LS}\approx-2(\meanUU\times\meanBB)\times\meanJJ
=-2(\meanJ_y\meanU_y)\meanBB.
\label{meanFFcLS2}
\end{equation}
Inspection of the data suggests that this is indeed a good approximation
and that therefore even in the simulation the normal component of
$\meanEMF\times\meanJJ$ is nearly vanishing on the boundary;
see \Fig{pcurhelflux}.
Both on the outer surface and on the equator $\meanFF_C^{\rm LS}$ is
such that it corresponds to a loss of negative current helicity.

In \Fig{pcurhelflux} we also show the small scale
current helicity fluxes on the two boundaries (fat lines).
There is a tendency for the difference between incoming
flux at the equator (dotted line) and outgoing fluxes at outer surface
(solid line) to cancel, but the net outgoing flux is again negative.
The flux for the total field is approximately four times larger than
what is accounted for by the Vishniac-Cho flux.
This might indicate that there is either another contribution to the
current helicity flux, or that the $\tau$ in the Vishniac-Cho flux
is underestimated.

\subsection{Application to the sun}

The purpose of this section is to put some real numbers into the
expression for the current helicity flux.
Our simulations have shown that a reasonable estimate
for the current helicity flux at the outer surface is
\EQ
\meanFF_C^{\rm SS}\approx30\,u_{\rm rms}k_{\rm f}B_0^2.
\EN
Applying this to the sun using $u_{\rm rms}\approx50\mpers$ for
the rms velocity in the deeper parts of the convection zone,
$k_{\rm f}\approx10^{-9}\cm^{-1}$ based on the inverse mixing length,
and $B_0\approx3\G$ for the mean field at the solar surface,
we have $\meanFF_C^{\rm SS}\approx10^{-3}\,\Gtwopers$.
The current helicity flux integrated over the northern hemisphere of the sun
is then $4\times10^{19}\G^2\cm^2\s^{-1}$.
Integrated over the 11 yr solar cycle we have $10^{28}\G^2\cm^2$.

\begin{figure}[t!]
\centering\includegraphics[width=0.4\textwidth]{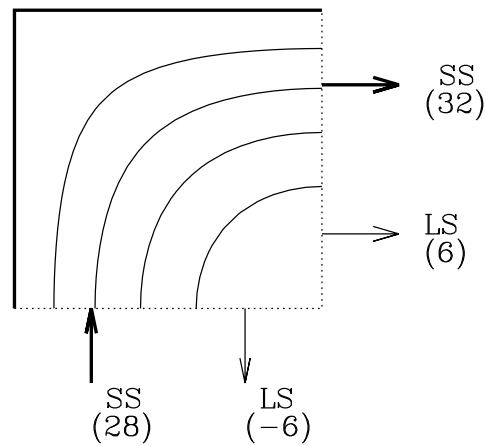}\caption{
Sketch illustrating the directions of large scale (LS) and small scale
(SS) negative current helicity fluxes and their approximate magnitudes
(in units of $\meanFF_0$).
Note that at the outer surface negative current helicity is ejected
both via small and large scale fields, while at the equator the
contributions from small and large scale fields have opposite sign.
The small scale losses at surface and equator partially cancel, giving
a net loss of negative current helicity of only about $4\,\meanFF_0$.
}\label{geom_flux}\end{figure}

For the sun only magnetic helicity fluxes have been determined.
As a rough estimate we may use
$\meanFF_H^{\rm SS}\approx k_{\rm f}^{-2}\meanFF_C^{\rm SS}$ for the
magnetic helicity flux.
Using the same estimate for $k_{\rm f}$ as above we obtain
about $10^{46}\Mx^2$ over the 11 yr solar cycle.
This is indeed comparable to the magnetic helicity fluxes estimated
by Berger \& Ruzmaikin (2000) and DeVore (2000).

We emphasize that, in the present context, ``small scale'' means
$2\pi/k_{\rm f}\approx60\Mm$, i.e.\ about one pressure scale height
at the bottom of the convection zone.
By contrast, ``large scale'' refers to length scales of the order of
several hundred Mm, which is typically beyond the scale captured in the
usual vector magnetograms.

\section{Conclusions}

The present simulations give a clear indication that the proposed set-up
with open boundaries and shear can alleviate the catastrophic quenching
problem in mean field dynamo theory.
Already in the absence of shear the resul\-ting $\alpha$ effect can be
larger when open boundary conditions are used, but there $\alpha$ still
decreases with increasing magnetic Reynolds number.
In the presence of shear, alleviated catastrophic quenching is
associated with a net loss of small scale current helicity.
At the equator, negative small scale current helicity flows into the
northern hemisphere, but there is an even larger negative small scale
current helicity flows ejected at the outer surface.
This results in a net loss of negative current helicity from the northern
hemisphere, and corresponds to an integrated magnetic helicity flux
of about $10^{46}\Mx^2$ over the 11 yr solar cycle.

Although our results are certainly encouraging, they must still be
considered preliminary.
First of all, we have still only considered a relatively limited range
of magnetic Reynolds numbers;
larger values are necessary before one can tell whether or not
an asymptotic $R_{\rm m}$ dependence still develops
for larger values of $R_{\rm m}$.
Secondly, turbulence simulations of a dynamo are necessary to
show that the electromotive force that here is interpreted in terms of
an $\alpha$ effect is indeed capable of generating large scale
magnetic field of the type shown in \Sec{Smeanfield}.
Of particular importance is the question whether the dynamo is oscillatory
(as expected from our mean field calculations)
and whether the cycle frequency is independent of $R_{\rm m}$.
Especially at large values of $R_{\rm m}$ the direct approach tends to
be advantageous compared to calculating $\alpha$ in the presence of an
imposed field, because the resulting $\alpha$ is always much more noisy
than the actual mean field obtained in a simulation (B01).

Another aspect to keep in mind is the fact that the correct boundary
conditions for the solar dynamo are certainly more complicated than the
vertical field condition adopted here.
There are good reasons to believe that the sun loses significant amounts
of magnetic and current helicity via coronal mass ejections (DeVore 2001,
D\'emoulin et al.\ 2002a,b, Gibson et al.\ 2002;
see also Blackman \& Brandenburg 2003).
The losses via via coronal mass ejections
are not easy to model within the present approach.
An obvious possibility is to include an outer layer
that resembles some important aspects of the solar corona (low density
and hence low plasma beta).
It is as yet unclear to which extent a spherical geometry is important.
The solar wind, for example, cannot be modeled in cartesian geometry,
and in some sense coronal mass ejections are just a particularly bursty
and localized manifestation of the solar wind.
It may therefore be worthwhile to consider the effects of boundaries
in global simulations.

\acknowledgements
We thank Eric Blackman, Detlef Elstner, and Kandu Subramanian
for extended discussions and detailed suggestions.
CS thanks NORDITA for hospitality.
The Danish Center for Scientific Computing is acknowledged
for granting time on the Linux cluster in Odense (Horseshoe).


\vfill\bigskip\noindent{\it
$ $Id: paper.tex,v 1.54 2004/07/27 19:35:18 brandenb Exp $ $}


\begin{thebibliography}{99}

\bibitem[]{}
Arlt, R., \& Brandenburg, A.\yana{2001}{380}{359}

\bibitem[]{}
Berger, M.\ygafd{1984}{30}{79}

\bibitem[]{}
Berger, M., \& Field, G. B.\yjfm{1984}{147}{133}

\bibitem[]{}
Berger, M., \& Ruzmaikin, A.\yjgr{2000}{105}{10481}

\bibitem[]{}
Blackman, E. G., \& Field, G. F.\yapj{2000a}{534}{984}

\bibitem[]{}
Blackman, E. G., \& Field, G. F.\ymn{2000b}{318}{724}

\bibitem[]{}
Blackman, E. G., \& Brandenburg, A.\yapj{2002}{579}{359}

\bibitem[]{}
Blackman, E. G., \& Brandenburg, A.\yapjl{2003}{584}{L99}

\bibitem[]{}
Brandenburg, A.\yapj{2001}{550}{824} (B01)

\bibitem[]{}
Brandenburg, A., \& Dobler, W.\yana{2001}{369}{329}

\bibitem[]{}
Brandenburg, A., \& Subramanian, K.\sjour{2004}{Phys.\ Rept.}
[arXiv:astro-ph/0405052]

\bibitem[]{}
Brandenburg, A., Nordlund, \AA., Pulkkinen, P.,
Stein, R.F., \& Tuominen, I.\yana{1990}{232}{277}

\bibitem[]{}
Brandenburg, A., Bigazzi, A., \& Subramanian, K.\ymn{2001}{325}{685}

\bibitem[]{}
Brandenburg, A., Dobler, W., \& Subramanian, K.\yan{2002}{323}{99}

\bibitem[]{}
Cattaneo F., \& Hughes D. W.\ypre{1996}{54}{R4532}

\bibitem[]{}
Choudhuri, A. R., Sch\"ussler, M., \& Dikpati, M.\yana{1995}{303}{L29}

\bibitem[]{}
DeVore, C. R.\yapj{2000}{539}{944}

\bibitem[]{}
D\'emoulin, P., Mandrini, C. H., van Driel-Gesztelyi, L.,
Lopez Fuentes, M. C., \& Aulanier, G.\ysph{2002}{207}{87}

\bibitem[]{}
D\'emoulin, P., Mandrini, C.~H., van Driel-Gesztelyi, L.,
Thompson, B.~J., Plunkett, S., Kov\'ari, Z., Aulanier, G.,
\& Young, A.\yapj{2002}{382}{650}

\bibitem[]{}
Dikpati, M. \& Gilman, P. A.\yapj{2001}{559}{428}

\bibitem[]{}
Durney, B. R.\ysph{1995}{166}{231}

\bibitem[]{}
Field, G. B., \& Blackman, E. G.\yapj{2002}{572}{685}

\bibitem[]{}
Gibson, S. E., Fletcher, L., Del Zanna, G., Pike, C. D., Mason, H. E.,
Mandrini, C. H., D\'emoulin, P., Gilbert, H., Burkepile, J., Holzer,
T., Alexander, D., Liu, Y., Nitta, N., Qiu, J., Schmieder, B., \&
Thompson, B. J.\yapj{2002}{574}{1021}

\bibitem[]{}
Ji, H.\yprl{1999}{83}{3198}

\bibitem[]{}
Kleeorin, N. I., \& Ruzmaikin, A. A.\yjour{1982}{Magnetohydro\-dynamics}{18}{116}

\bibitem[]{}
Kleeorin, N. I, Rogachevskii, I., \& Ruzmaikin, A.\yana{1995}{297}{159}

\bibitem[]{}
Kleeorin, N. I, Moss, D., Rogachevskii, I., \& Sokoloff, D.\yana{2000}{361}{L5}

\bibitem[]{}
Kleeorin, N. I, Moss, D., Rogachevskii, I., \& Sokoloff, D.\yana{2002}{387}{453}

\bibitem[]{}
Kleeorin, N. I, Moss, D., Rogachevskii, I., \& Sokoloff, D.\yana{2003}{400}{9}

\bibitem[]{}
Krause, F., \& R\"adler, K.-H.\ybook{1980}
{Mean-Field Magneto\-hydrodynamics and Dynamo Theory}
{Akademie-Verlag, Berlin; also Pergamon Press, Oxford}

\bibitem[]{}
Moffatt, H. K.\ybook{1978}
{Magnetic Field Generation in Electrically Conducting Fluids}
{Cambridge University Press, Cambridge}

\bibitem[]{}
Ossendrijver, M., Stix, M., \& Brandenburg, A.\yana{2001}{376}{713}

\bibitem[]{}
Parker, E. N.\ybook{1979}{Cosmical Magnetic Fields}{Clarendon Press, Oxford}

\bibitem[]{}
Pouquet, A., Frisch, U., \& L\'eorat, J.\yjfm{1976}{77}{321}

\bibitem[]{}
Roberts, P. H.\yptrs{1972}{A272}{663}

\bibitem[]{}
Roberts, P. H., \& Stix, M.\yana{1972}{18}{453}

\bibitem[]{}
Rogachevskii, I. \& Kleeorin, N.\ypre{2003}{68}{036301}

\bibitem[]{}
Seehafer, N.\ypre{1996}{53}{1283}

\bibitem[]{}
Subramanian, K.\yjour{2002}{Bull.\ Astr.\ Soc.\ India}{30}{715}

\bibitem[]{}
Subramanian, K., \& Brandenburg, A. 2004 (in preparation)

\bibitem[]{}
Vishniac, E. T., \& Cho, J.\yapj{2001}{550}{752}

\bibitem[]{}
Zeldovich, Ya. B., Ruzmaikin, A. A., Sokoloff, D. D.\ybook{1983}
{Magnetic fields in astrophysics}{Gordon \& Breach, New York}

\end{thebibliography}
\end{document}